# Nature of Field Galaxies in the (z ≤ 0.3) Universe from the CFRS Survey

L. Tresse, C. Rola, F. Hammer, G. Stasińska

# NATURE OF FIELD GALAXIES IN THE (Z ≤ 0.3) UNIVERSE FROM THE CFRS[†] SURVEY

L. TRESSE[1,*], C. ROLA[1,2], F. HAMMER[1,*] and G. STASIŃSKA[1]
[1] *D.A.E.C., Observatoire de Paris-Meudon,*
*92195 Meudon Principal Cedex, France*
[2] *Centro de Astrofísica da Universidade do Porto,*
*Rua do Campo Alegre, 823, 4100 Porto, Portugal*

## ABSTRACT

We present the first detailed spectroscopic study of field galaxies up to $z = 0.3$ from the I-magnitude limited CFRS[†] sample. In this complete sample, we find that about half of the objects are blue emission-line galaxies. Using line ratio diagrams and photoionization models, we show that in roughly half of these emission-line galaxies, ionization by hot stars cannot account for the observed line ratios. Their spectra have properties intermediate between Seyfert 2 galaxies and low-ionization nuclear emission-line regions (LINERs). Their number density (about 20%) relative to the total number of galaxies up to $z = 0.3$, is much larger than the number density (2%) of active galactic nuclei (AGN) in the local Universe. Thus, the observed excess of blue galaxies in counts down to $B = 24$ cannot be mainly due to starburst galaxies, as was previously thought. This result can shed a new light on the nature of the overabundant population of blue galaxies at low redshifts.

## 1. The Sample of Field Galaxies up to z = 0.3

### 1.1. A Complete Sample

The CFRS[†] sample ($z \leq 1$) presents several useful aspects to study field galaxies at low redshifts. Its I-band magnitude-limited photometric catalog ($17.50 \leq I_{AB} \leq 22.50$) allows to select field galaxies from their old stellar population, since the observed light is emitted at wavelenghs greater than 4000 Å. Thus, the sample is less sensitive to star forming galaxies, which dominate the B-band selected samples. For the spectroscopic observations, the sources are picked up using an isophotal magnitude scheme, which avoids a bias in favor of compact sources. The ($4500 - 8500$Å) spectral range allows to see the Balmer line $H_\alpha$ up to $z \simeq 0.3$. When this line and $H_\beta$ are in emission, we are able to correct the emission-line galaxies for reddening using the Balmer decrement given by the ratio $H_\alpha/H_\beta$. Morever, there is no bias in favor of strong emission-line objects (with high equivalent width of $H_\alpha$) since for all objects $EW(H_\alpha)$ is much lower than the band pass of the I filter. The completeness of the CFRS sample reaches about 85%. The larger part of the incompleteness might occur at high redshifts (see D. Crampton's contribution). Some incompleteness is also expected at low redshifts. We have used a filter which cuts wavelengh below 4500 Å which makes the break at 4000Å invisible for galaxies at redshift less than $z \simeq 0.1$. This brings difficulties in attributing a redshift for some absorption-line spectra of low-$z$ objects. Nevertheless, the subsample of field galaxies up to $z = 0.3$, is a quite good representation of the field population at low redshifts.

---

[†] The Canada-France Redshift Survey's team is composed by D. Crampton, F. Hammer, O. Le Fèvre, S. Lilly and L. Tresse, see proceedings by S. Lilly and D. Crampton.

[*] Visiting observer with the Canada-France-Hawaii Telescope, operated by the NRC of Canada, the CNRS of France and the University of Hawaii.

*1.2. Goals*

The discovery that the blue excess in the deep counts of the field galaxy population is due to faint blue galaxies at low redshifts (Broadhurst *et al* 1988), has yield a puzzling problem. The recent deep spectroscopic surveys (see Colless's and Cowie's contributions, Lilly 1993, Tresse *et al* 1993) are consistent with the no-evolution models despite a signifiant excess in number to $B = 24$, relatively to expectations from the local density. Results from the CFRS data (see S. Lilly's contribution, Hammer 1994, Le Fèvre 1994) show no evolution for galaxies redder than Sbc, while the blue ones present about one magnitude of luminosity evolution at high redshifts ($L^*$ galaxies) and a significant number excess at low redshifts (sub-$L^*$ galaxies). Then what has happened to these faint galaxies? Several models have been suggested: the fading (Babul & Rees 1992) and/or the merging of low-luminosity galaxies (see Guiderdoni and Broadhurst's communications) or the existence of an unrecognized population of objects (Cowie 1991). The faint blue galaxies were often classified as starburst galaxies, but no detailed studies have been made to identify the real nature of their ionization sources. A first qualitative approach about the nature of these galaxies through their physical properties is presented here. The aims of our analysis are to classify galaxies up to $z = 0.3$ and to investigate which kind of objects would be responsible for the blue number excess.

## 2. Classification of the Sources

The complete subsample up to $z = 0.3$ is composed by 139 field galaxies having a well-defined redshift (with a note> 1, see S. Lilly's contribution). We classify their spectra in three categories using the following criteria. The first category is composed by 36 sources (26%) having a spectrum with $H_\alpha$ in emission and $H_\beta$ in absorption. The second category contains 80 sources (57%) exhibing four emission lines or more, $H_\alpha$, $H_\beta$, [OIII]$\lambda$4959, [OIII]$\lambda$5007, and including generally in the observed spectral range [OII]$\lambda$3727 and/or [SII]$\lambda$6725. These spectra have a blue continuum and narrow emission lines. The third category contains 23 sources (17%) having a spectrum with only absorption lines. They present usually a red continuum (figure 1). For our analysis, we deal with the second category of spectra since the quality of the spectra and their emission lines allow us to study it in detail. They represent more than half of the field population up to $z = 0.3$ and they correspond to the faint blue galaxy population.

## 3. Study of the Emission-Line Spectra

## 3.1. Analysis of the Data and Line Ratio Diagrams

With the package SPLOT under the software IRAF and the MEASURE tool (code developed by D. Pelat (Meudon), an extension of which is presented in Rola & Pelat 1994), integrated intensities and errors at one sigma are computed for each emission line of spectra belonging to the second category. When [NII]$\lambda$6583 is seen, we deblend H$_\alpha$ from [NII]$\lambda$6583 and [NII]$\lambda$6548 with the deblend utility under the SPLOT package. The intensity line ratios [OIII]$\lambda$5007/H$_\beta$, [OII]$\lambda$3727/H$_\beta$ and [SII]$\lambda$6725/H$_\alpha$ are calculated and dereddenned. The emission-line [OI]$\lambda$6300 is usually not detected above the noise of the continuum. The reddening constant is computed using the Balmer decrement H$_\alpha$/H$_\beta$ equal to 2.86 (case B, $T = 10\ 000K$, $n = 100\ cm^{-3}$, Osterbrock 1989) and the Seaton reddening law (Seaton 1979). Since the 5 fields of the CFRS are at high galactic latitude ($|b_{II}| \geq 50^\circ$), the reddening is assumed to be intrinsic to the observed galaxies.

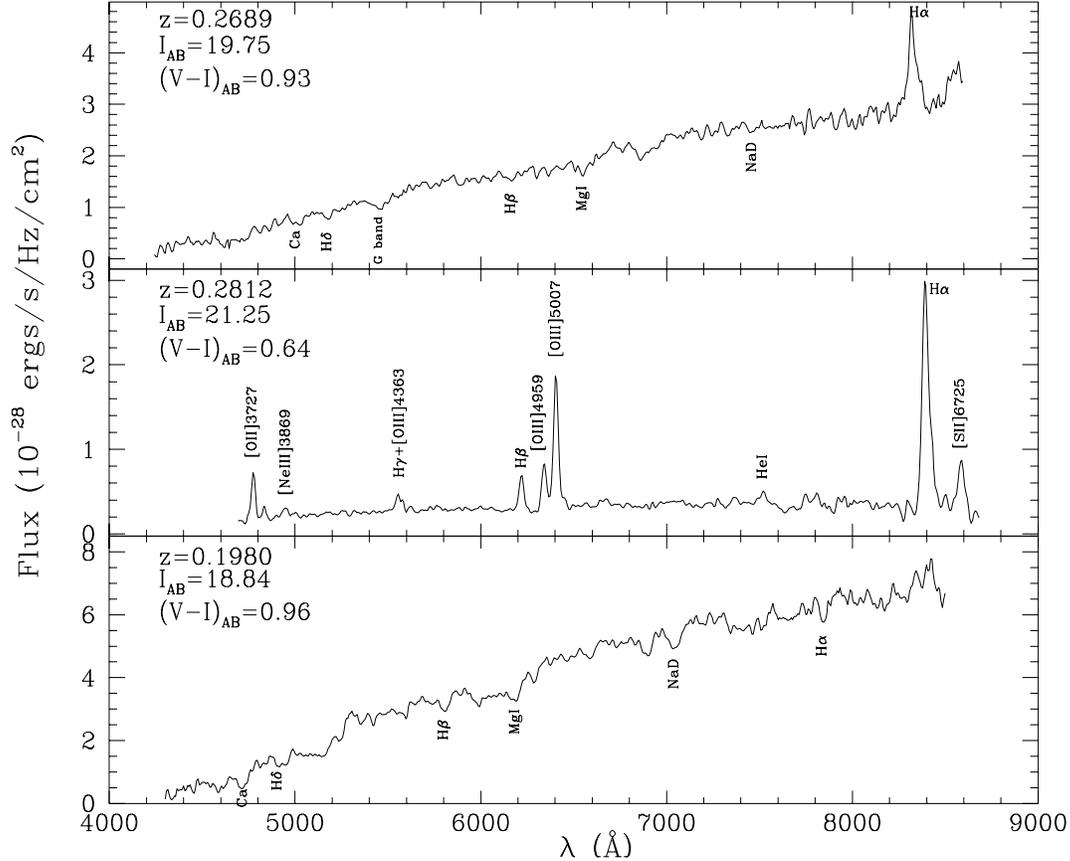

Fig. 1. Classes of Spectra. At the top: spectra with H$_\alpha$ in emission and H$_\beta$ in absorption (36 spectra), in the middle: spectra with H$_\alpha$ and H$_\beta$ in emission (80 spectra), at the bottom: spectra with H$_\alpha$ and H$_\beta$ in absorption (23 spectra).

To identify the nature of the ionization source, our line ratios have been plotted in the [OIII]$\lambda$5007/H$_\beta$ versus [SII]$\lambda$6725/H$_\alpha$ diagnostic diagram (Veilleux & Osterbrock

1987) where HII region-like objects and AGN lie in two distinct areas. Our most puzzling result is that a significant fraction of sources fall in the AGN area and in particular, in the Seyfert 2 galaxies area (figure 2). Other sources fall in the area of HII region-like objects with high ionization parameter, and often in the starburst galaxies area (between the HII region-like objects area and Seyfert 2 galaxies area). The advantage of this diagram is that the intensity line ratios involve lines with small wavelength separation, minimizing uncertainties introduced in relative flux calibration and reddening correction.

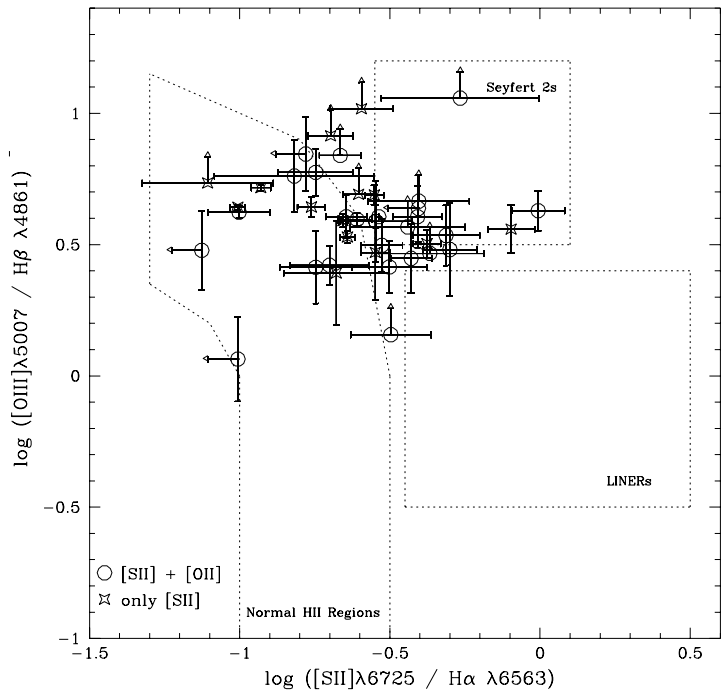

Fig. 2. Reddening-corrected [OIII]$\lambda$5007/H$_\beta$ vs. [SII]$\lambda$6725/H$_\alpha$ intensity ratios. Error bars are at 1$\sigma$ level. Circle symbols represent spectra with [OII]$\lambda$3727 and [SII]$\lambda$6725 inside the observed spectral range. Star symbols represent spectra with [OII]$\lambda$3727 outside the spectral range. Dashed curves delineate typical areas corresponding to normal HII regions, Seyfert 2 galaxies and LINERs from Filippenko & Terlevich (1992).

To go further with this analysis, we have also considered the [OIII]$\lambda$5007/H$_\beta$ versus [OII]$\lambda$3727/H$_\beta$ diagnostic diagram. This diagram depends on the reddening. The sources have a strong [OII]$\lambda$3727 - log([OII]$\lambda$3727/H$_\beta$)$\geq$ 0 - and many are far of the HII main sequence computed by McCall (1985). Most of the sources belonging to the low-ionization Seyfert 2 galaxies area from the previous diagnostic diagram have the strongest [OII]$\lambda$3727/H$_\beta$ line ratios (figure 4).

From this first analysis, we can conclude that a part of these emission-line galaxies have spectra similar to Seyfert 2 galaxies (figure 3) and cannot be HII region-like objects or starburst galaxies. This is in contradiction with the idea prevaling so far that faint blue galaxies are mostly startburst galaxies. The latter conclusion was mostly derived from the [OII]λ3727 equivalent width. But it is only by observing several emission lines and considering appropriate line ratios that one can safely classify an emission-line object.

### 3.2. Comparison with Photoionization Models

To find the limits of photoionization by massive stars in the diagnostic diagrams cited above, we have used the photoionization code PHOTO (Stasińska 1990), and constructed a grid of spherically symmetric HII region models. The density was taken constant ($n_H = 10$ $cm^{-3}$) and the values of the ionization parameter $U = Q_H/(4\pi R^2 n_H c)$ (where $Q_H$ is the number of hydrogen ionizing photons, $R$ is the Strömgren radius and $c$ is the light speed) were varied between 1 and 2 $10^{-6}$. For simplicity, we have considered the ionizing source as a single star with an effective temperature, $T_{eff}$, which was varied from 3 $10^4$ to 6 $10^4$K. For the distribution of the ionizing radiation, we have used the $\log g = 5$ Kurucz model atmospheres (Kurucz 1992) with an abundance consistent with the one in the nebula. Several abundance sets were considered: 2 $Z_\odot$, $Z_\odot$, 0.25 $Z_\odot$, 0.1 $Z_\odot$ (in each case, the abundances of the elements relative to oxygen were chosen accordingly to the expressions given by McGaugh 1991). Figure 4 shows the loci in the two diagnostic diagrams, for the models which give the upper maximum boundary for photoionization by main sequence OB stars. These models have $n_H = 10$ $cm^{-3}$, $T_{eff} = 60\ 000K$ (according to the Maeder 1990 models this effective temperature corresponds to a stellar mass of 120 $M_\odot$) and their corresponding metallicities are stated in the figure caption. The thick line represents the curve empirically determined by Veilleux & Osterbrock (1987) from observational data, separating AGN from HII region-like objects. Our sample of emission-line galaxies is superimposed in both diagrams. These plots confirm that a significant fraction of the objects lies above and to the right of the boundary defined by our models, and therefore cannot be ionized by main sequence massive stars.

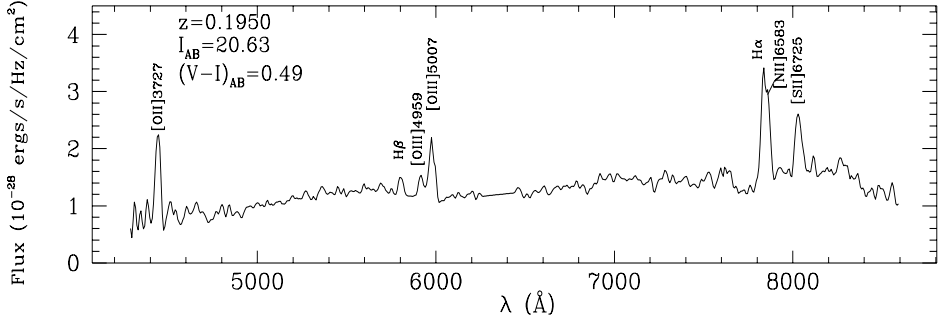

Fig. 3. Example of an object with a Seyfert 2-like spectrum.

We found it useful to estimate the blackbody effective temperatures which would account for the galaxies of our sample that do not have HII region-like spectra. In about half of these objects, temperature over 100 000 K are necessary to explain the position in the [OIII]λ5007/H$_\beta$ versus [OII]λ3727/H$_\beta$ plot. This is too hot for evolved massive stars (Leitherer et al 1992). Morever, the warmer stage is too short to account for such a large proportion of non-HII region-like spectra in our sample. Ionization by post-AGB stars from an old stellar population (Binette *et al* 1994) is ruled out since it would produce much smaller emission-line equivalent width than observed in our galaxies (about 30-40 Å) and red colors instead of blue. We are thus left with the conclusion that many galaxies in our sample are genuine active galaxies.

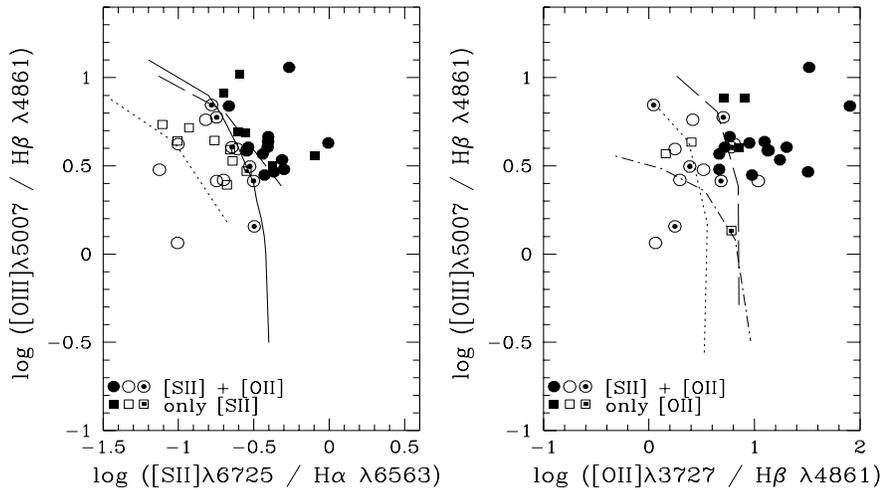

Fig. 4. Lines ratio diagrams. Round symbols represent spectra with both [SII]$\lambda$6725 and [OII]$\lambda$3727 in the observed spectral range. Square symbols represent spectra with only [SII]$\lambda$6725 or only [OII]$\lambda$3725 inside the spectral range. Black symbols are data that are in the AGN area at more than $1\sigma$ from the normal HII regions area (see fig. 2). Open symbols are data in the normal HII regions area. White and black symbols are data in the intermediate area (solid line representing the Veilleux & Osterbrock's separation). Models of the loci for photoionization by OB stars with $T_*$=60 000 K, are represented by long dashed lines for Z=0.25 $Z_\odot$, by dot lines for Z=0.1 $Z_\odot$, and by dot-dashed line for Z=$Z_\odot$.

## 4. Statistics

For the 80 spectra of the second category, 45 spectra have neither [OII]$\lambda$3727 nor [SII]$\lambda$6725 in the spectral range, a sky emission line affects one emission line for 4 spectra and one or several emission lines are affected by instrumental defects (zero order) for 4 other spectra. Then for 27 spectra, we were able to do our analysis with both diagnostic diagrams. We assume that these 27 spectra are representative of all the 80 spectra. Amongst them, 13 are placed in the AGN area in both diagnostic diagrams. This represents 28% of all ($I_{AB} < 22.5$) the field galaxies of our sample whose redshift has been measured and is smaller than 0.3. Even if two third of the galaxies with unknown redshift had $z \leq 0.3$, the fraction of AGN amongst field galaxies up to $z = 0.3$ would be 15%. Considering the nature of the incompleteness of the CFRS survey, a proportion of about 20 % is a more realistic estimate. These numbers have to be compared to the much smaller fraction (2%) of low-luminosity AGN sources found locally in magnitude-limited spectroscopic sample (Huchra & Burg 1993).

## 5. Conclusion

We have analysed the nature of emission-line galaxies from the complete CFRS sample up to $z = 0.3$ using diagnostic diagrams and photoionization models. The faint blue galaxies (bluer than Sbc) represent almost half of the field galaxy population up to this redshift. Half of them have spectroscopic properties similar to low-ionization Seyfert 2 galaxies. Their average characteristics are $\bar{z} \simeq 0.2$, $\overline{M}_{B_{AB}} \simeq -18.4$ ($H_0 = 50$ km s$^{-1}$ Mpc$^{-1}$), $\overline{(V-I)}_{AB} \simeq 0.45$. The images of the CFRS sample show that some of these galaxies are spirals edge-on, other are compact sources or irregular sources. Morphological studies will be done in the next future to deconvolve the nucleus emission from the blue starlight of these galaxies.

The excess in counts down to B=24 of faint blue galaxies was previously attributed to starburst galaxies. Our work shows that about 20% of all the field galaxies at $z \leq 0.3$ are narrow line AGN. This strongly modifies the current representation of the Universe at low redshifts.

A full version of this work will be submitted to MNRAS.

## Acknowledgements

We thank David Crampton, Olivier Le Fèvre and Simon Lilly for their information, advice and help. We thank Catherine Boisson, Marshall MacCall, Jean-René Roy, and Roberto Terlevich for helpful discussions. L. T. thank the EC Human Capital and Mobility Support for financial support to attend this Conference.